# Comment on: "Nonlinear quantum effects in electromagnetic radiation of a vortex electron"

Aviv Karnieli,[1] Roei Remez,[1] Ido Kaminer[2] and Ady Arie[1]

[1]School of Electrical Engineering, Fleischman Faculty of Engineering, Tel Aviv University, Tel Aviv, Israel
[2]Department of Electrical Engineering, Technion–Israel Institute of Technology, Haifa 32000, Israel

**Abstract**

This comment on the *Phys. Rev. A* paper "Nonlinear quantum effects in electromagnetic radiation of a vortex electron" by Karlovets and Pupasov-Maximov [*Phys. Rev. A* **103**, 12214 (2021)] addresses their criticism of the combined experimental and theoretical study "Observing the quantum wave nature of free electrons through spontaneous emission" by Remez et al, published in *Phys. Rev. Lett.* [*Phys. Rev. Lett.* **123**, 060401 (2019)]. We show, by means of simple optical arguments as well as numerical simulations, that the criticism raised by Karlovets and Pupasov-Maximov regarding the experimental regime reported by Remez et al is false. Further, we discuss a necessary clarification for the theoretical derivations presented by Karlovets and Pupasov-Maximov, as they only hold for a certain experimental situation where the final state of the emitting electron is observed in coincidence with the emitted photon – which is not the common scenario in cathodoluminescence. Upon lifting the concerns regarding the experimental regime reported by Remez et al, and explicitly clarifying the electron post-selection, we believe that the paper by Karlovets and Pupasov-Maximov may constitute a valuable contribution to the problem of spontaneous emission by shaped electron wavefunctions, as it presents new expressions for the emission rates beyond the ubiquitous paraxial approximation.

**Introduction**

The point of controversy is the interpretation of the free-electron wavefunction in the Smith-Purcell effect[2]. In particular, the new *Phys. Rev. A* paper argues against the validity of the experimental parameters and conclusions reported in Ref[2]. We identify two incorrect assumptions in the *Phys. Rev. A* paper that leads to (i) a faulty claim against the validity of the far-field measurements in Ref[2] and (ii) inaccurate modeling of the Smith-Purcell effect in the experiment, and any free-electron radiation process. These mistakes explain the conflicting theoretical results between Ref[1] and Ref[2], and the inability of Ref[1] to explain the experiment in Ref[2]. Below, we explain the necessary physical assumptions in two ways: using simple optical considerations, and numerical simulations, both clarifying why the conclusions in Ref[2] hold.

**1. Validity of the measurements and conclusions in Remez et al[2]**

1a. Semiclassical and quantum theories of spontaneous emission by free electrons

Before we delve into the quantitative estimate, we make a short reminder about the context of the study reported by Remez et al[2]. The experiment and complementing theory aimed to decide between two possible interpretations of the role of a free electron's wavefunction in the process of spontaneous emission from that electron: (1) **the semiclassical interpretation**: the electron emits light coherently as a smeared-out charge density given by $\rho(\mathbf{r}) = e|\psi(\mathbf{r})|^2$; (2) **the quantum interpretation**: the electron effectively "collapses" to a point $\mathbf{r}$ upon emission with probability $|\psi(\mathbf{r})|^2$, and then emits light incoherently from different points (or, localized regions of size much smaller than the emitted wavelength). In Ref[2], it was found that both experiment and QED-based theory agree with the second interpretation.

Ref[1] claims that a semiclassical interpretation could still explain the physics of the experiment reported in Ref[2] – and that its results cannot reject the semiclassical interpretation. The main reason cited by the authors of Ref[1] is that the experiment in Ref[2] was performed in near-field conditions. It is the purpose of the following section to show that **the experiment reported in Ref[2] rejects the semiclassical theory** and that it was performed in the far-field as claimed in the paper[2]. We explicitly show that the semiclassical and quantum interpretations predict dramatically different radiation patterns. To do this, below we first consider the predictions made by the semiclassical interpretation – removing the ambiguity suggested by Ref[1] – and then explain how, based on the measurements reported in Ref[2], the semiclassical interpretation can be rejected in favor of the quantum interpretation.

1b. Fraunhofer diffraction condition of light emitted by partially coherent electron beams

Karlovets and Pupasov-Maximov[1] claim that the experimental results reported by us in Ref[2] do not agree with their theoretical prediction because the radiation pattern was presumably not measured in the far-field. Based on this argument, they claim that the divergence observed in the radiation can still be accounted for via a semiclassical near-field effect. Put in other words, the authors argue that the Fraunhofer diffraction condition $r \gg d^2/\lambda$ (Eq. 6 in their paper) is not met in the measurements reported in Ref[2] ($r$ is the observation distance, $d$ the aperture diameter and $\lambda$ the optical wavelength). The reason they provide is that the width of the *entire* electron beam in the experiment in Ref[2] is in the range of $d = w_{\text{beam}} = 0.3$ mm to 2 mm, implying that in order to be in the far-field, $r$ must exceed 0.15 m and 6.7 m, respectively, wherein the radiation was observed at a distance of approximately $r = 0.25$ m.

Here we explain the mistake in this argument. The relevant parameter for calculating the interference of light emitted from a current source is the transverse coherence length – denoted by $l_c$ – of the source, and not its total incoherent width ($w_{\text{beam}}$) as was assumed in the calculations of Ref[1] when claiming against the measurements in Ref[2]. The incoherent width can be thought of as a statistical distribution, where each arriving electron appears at a different location. Each single electron therefore consists of an independent wavefunction coherent only with itself, on a transverse length scale $l_c$. The coherent radiation emitted by each single electron, therefore, could only depend on a scale of up to $l_c$, and not $w_{\text{beam}}$.

This means that the correct effective aperture $d$ from which the light is emitted in the case of a semi-classical calculation (as Ref[1] employs) would be on the order of the transverse coherence length, i.e., $d = l_c$ and *not* $d = w_{\text{beam}}$, such that the Fraunhofer condition reads

$$r \gg l_c^2/\lambda.$$

In the experiment reported in Ref[2], the transverse coherence length of the electron was measured to be $l_c = 5\mu m$ for $w_{\text{beam}} = 0.3$mm and $l_c = 33\mu m$ for $w_{\text{beam}} = 2$mm. Consequently, the Fraunhofer condition is easily met by the measurements, giving $r \gg 50\mu m$ and $r \gg 2.2$ mm, respectively, where the distance in the experiment reported in Ref[2] was $r = 0.25$m. This implies that Ref[2] safely meets the far-field condition. We further note that to avoid **geometrical** influence on the measurement (e.g., a parallax) one needs to ensure that the region of the different sources (i.e., the beam width $w_{\text{beam}}$) spans a small angle in the observation plane, which is also easily met by the experiment since $w_{\text{beam}}/r \ll 1$.

We complement these arguments with a numerical FDTD simulation of Smith-Purcell emission by free electrons passing near a metallic grating. For numerical convergence purposes, we scaled-down the problem to SEM energies ($E = 30$keV) such that the grating length was $L = 4$μm with a periodicity of $\Lambda = 200$nm for visible emission at $\lambda = 600$nm for $\theta = 90°$, and an incoherent beam width of $w_{\text{beam}} = 20$μm. We varied the coherence length $l_c$ between $0.2$μm, $1$μm and $4$μm, adding **incoherently** the contributions to the emitted light from different parts of the electron beam (i.e., adding incoherently 100 simulations with $l_c = 0.2$μm, 20 simulations with $l_c = 1$μm, etc.) – see Fig. 1. We propagated the near-field to a distance $r = 100$μm. This distance is **larger** than the Fraunhofer distance associated with the coherence lengths ($r \gg l_c^2/\lambda$, with $l_c^2/\lambda$ found to be at most $27$μm for $l_c = 4$μm), while being **smaller** than the Fraunhofer distance associated with the incoherent beam width $w_{\text{beam}}$, i.e. $r \ll w_{\text{beam}}^2/\lambda = 667$μm, as falsely argued by Ref[1]. We find that, as expected, the divergence varies dramatically according to the coherence length $l_c$, even though the beam width $w_{\text{beam}}$ stayed the same.

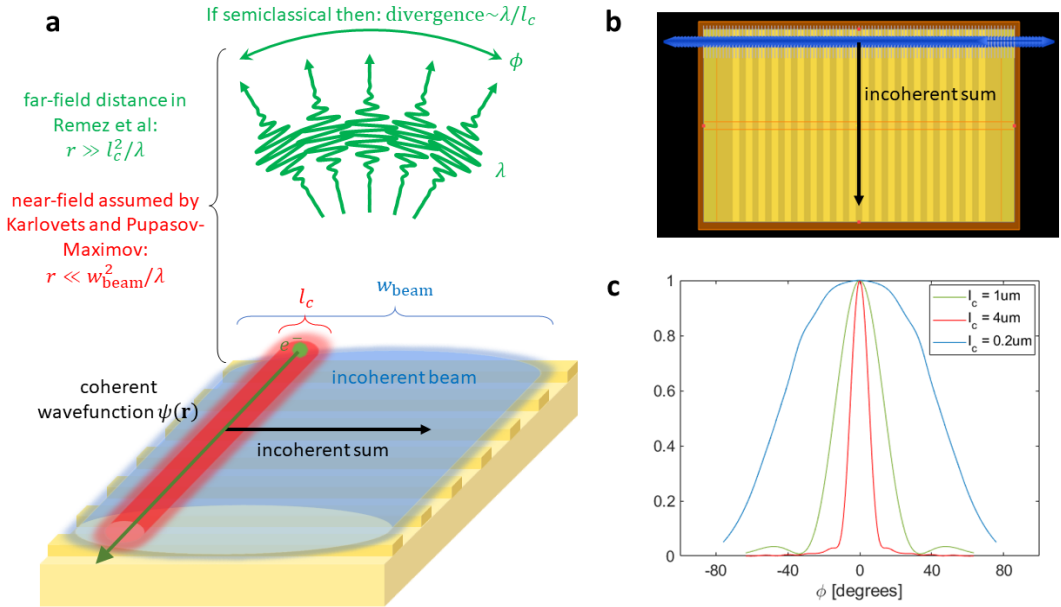

Figure 1. **Numerical FDTD simulations of Smith-Purcell radiation by partially coherent electron beams. a** Illustration of the Smith-Purcell emission from a partially coherent source – under the semiclassical (charge density) interpretation. The incoherent electron beam width $w_{\text{beam}}$ is much larger than the wavefunction coherence length $l_c$. Light with wavelength $\lambda$ is emitted coherently from each aperture $l_c$, and the radiation is summed incoherently over the beam width. A far-field observer located at $r \gg l_c^2/\lambda$ measures the light divergence along the azimuthal angle $\phi$. If the light emission is semiclassical (i.e., light is emitted coherently from a smeared-out charge density), the observer should see a divergence that scales with $\lambda/l_c$. In the experiment reported in Ref[2] this logic was used to rule out the classical interpretation in favor of a quantum probabilistic point-charge model, as no change in the divergence was observed for a corresponding change in $l_c$. **b** FDTD model used for the semiclassical simulations. For each simulation, a coherence length was chosen and the emitted light was summed incoherently from different parts of the beam width. **c** The angular spectrum along the azimuthal angle as predicted by the semiclassical interpretation for different coherence lengths. The semiclassical theory predicts a considerable change in the angular divergence when $l_c$ is varied. This result was obtained upon propagating the near-field to a distance of $r = 100$μm away from the source, satisfying $r \gg l_c^2/\lambda$ but at the same time ensuring $r \ll w_{\text{beam}}^2/\lambda$ as was in the experiment[2]. In contrast, Karlovets and Pupasov-Maximov[1] mistakenly considered the Fraunhofer condition to be $r \gg w_{\text{beam}}^2/\lambda$, suggesting that the measurements in Ref[2] were taken in the near-field. They further claim that in this case, the radiation angular divergence should be wide for all values of $l_c$ due to a near-field effect. Our simulations explicitly show that this claim is wrong, and that the radiation patterns strongly depend on $l_c$ even in the regime which Karlovets and Pupasov-Maximov[1] considered as the near-field.

1c. Ruling out the semiclassical theory

The calculation of spontaneous emission under the semiclassical interpretation considers a coherent interference from an extended source defined by $\rho(\mathbf{r}) = e|\psi(\mathbf{r})|^2$, where the width of $\psi(\mathbf{r})$ is of the order of $l_c$ – thus defining the effective aperture $d = l_c$. At the same time, there is also an incoherent summation of the field intensities emitted from different parts of the electron beam, as the incoherent width of the electron beam $w_{\text{beam}}$ is much wider than $l_c$. Under this interpretation, a change in $l_c$ (between $5\mu m$ and $33\mu m$) should, in the far-field, lead to a respective change in the azimuthal divergence $\Delta\phi \sim \lambda/l_c$. Should no such change be observed, the semiclassical interpretation can be ruled out in favor of the incoherent point-like emission, quantum interpretation, predicted by quantum electrodynamics.

The experiment in Ref[2] did not observe any change in the measured divergence, even though the coherence length $l_c$ was varied by more than 6 times. This evidence therefore implies that **we can reject the semiclassical interpretation in favor of the quantum interpretation**, suggesting an invariant divergence.

## 2. Modelling of spontaneous emission by free electrons: the role of electron post-selection

The key point is that Ref[1] derives emission rates that depend on the final electron quantum state. This derivation corresponds to a post-selection of the electron (or coincidence measurement of the electron and photon), which does not happen in the experimental setup of Ref[2]. Although such a post-selection is experimentally feasible[3], this is not the common situation in free-electron radiation (e.g. see literature on cathodoluminescence[4,5]), where only the light is measured. In contradiction with the prediction in Ref[1], there is now strong evidence[2,6,7] that in the case of regular free-electron radiation with no electron post-selection, the radiation does not normally depend on the initial electron wavefunction, due to the underlying electron-photon entanglement leading to decoherence[8,9]. Fig. 2 illustrates this principle by showing that for light emitted by a coherent wavefunction, interference is washed out if the electron final state is not observed (i.e., no post selection of the final state is made).

There are possible exceptions to this rule in special situations in which the electron wavefunction does matter in spontaneous emission processes: for example, when an external potential mediates scattering of initial electron states into a common final state[10], or when quantum recoil corrections[7,11] are dominant.

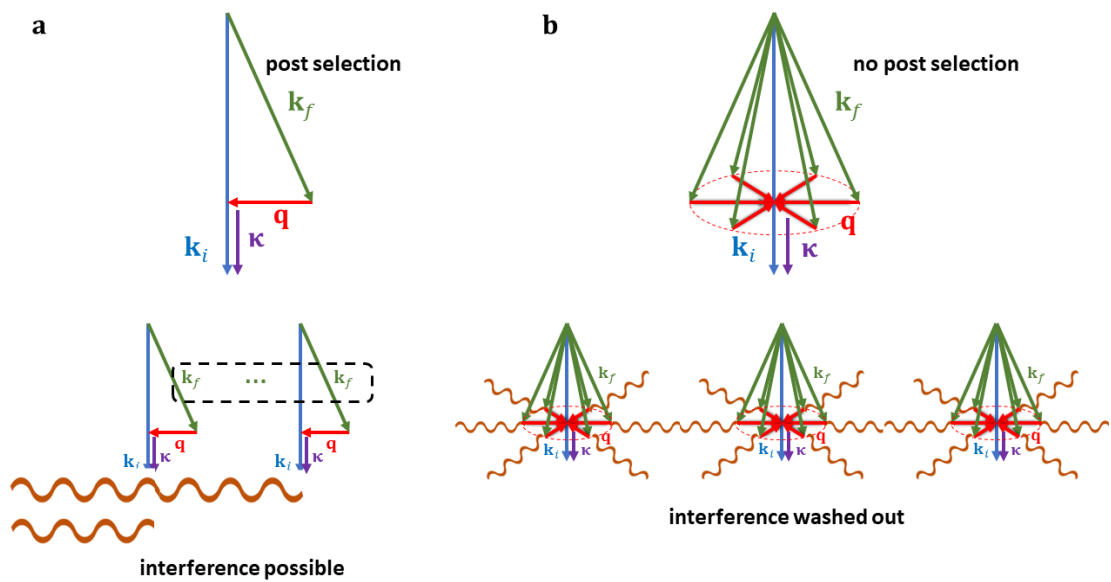

Figure 2. **Illustration of light emission by post-selected and traced-out electron final states. a** Free-electron emission from an initial electron momentum state $\mathbf{k}_i$ that scatters due to the emission into a post-selected final state $\mathbf{k}_f$, where the emitted light has a wavevector $\mathbf{q}$, and the emission is mediated by the grating momentum $\mathbf{\kappa}$. Momentum and energy conservation impose entanglement between the electron and photon. Post-selecting electron final states allows the light emitted from different initial states of the wavefunction to interfere, making the emission wavefunction-dependent. **b** when the electron final state is not measured, the entanglement between all possible final states and all possible photonic modes leads to quantum decoherence. In this case, the emission is wavefunction-independent, i.e., light interference from different initial states of the electron wavefunction is washed out.

## 3. Conclusions

We lift the concerns regarding the experimental parameters used in Ref[2], raised by the authors of Ref[1]. We show both by simple optical arguments as well as numerical simulations, that the measurements reported in Ref[2] took place in the far-field with respect to the electron transverse coherence length. From these measurements[2], we conclude that it is possible to reject the semiclassical theory in favor of the quantum theory.

We further suggest the authors clarify that the derivations following Eq. 10 in their paper[1] correspond to an experimental situation where the electron final state (after emission of light) is post-selected (or measured in coincidence with the light). This is a ***crucial detail***, since in this case the radiation be dependent on the electron initial wavefunction[7,12], but **the physical scenario is completely different**. The common situation for measuring cathodoluminescence involves only measuring the light (as was done in the experiment of Ref[2]), which corresponds to **tracing out** the electron final states[13]. This last step is what allows for the incoherent summation over many "point electrons" emitting light – the quantum interpretation – to be correct.

Upon lifting the concerns of Ref[1], and explicitly noting the key requirement for electron post-selection, we believe that Ref[1] may constitute a valuable contribution to the problem of spontaneous emission by shaped electron wavefunctions. Specifically, the novelty of Ref[1] is in deriving new expressions for the emission rates beyond the ubiquitous paraxial approximation and in showing that such rates could depend on the wavefunction upon post-selection.